# Can a half-metallic zincblende-type structure be stabilized via epitaxy?

Yu-Jun Zhao and Alex Zunger
*National Renewable Energy Laboratory, Golden, Colorado 80401*

The need for spin-injectors having the same zincblende-type crystal structure as conventional semiconductor substrates has created significant interests in theoretical predictions of possible metastable "half-metallic" zincblende ferromagnets, which are normally more stable in other structure-types, e.g., NiAs. Such predictions were based in the past on differences $\Delta_{\rm bulk}$ in the total-energies of the respective *bulk* crystal forms (zincblende and NiAs). We show here that the appropriate criterion is comparing difference $\Delta_{\rm epi}(a_s)$ in *epitaxial* total-energies. This reveals that even if $\Delta_{\rm bulk}$ is small, still for MnAs, CrSb, CrAs, CrTe, $\Delta_{\rm epi}(a_s) > 0$ for all substrate lattice constant $a_s$, so the zincblende phase is not stabilized. For CrS we find $\Delta_{\rm epi}(a_s) < 0$, but the system is antiferromagnetic, thus not half-metallic. Finally, zincblende CrSe is predicted to be epitaxially stable for $a_s > 6.2$ Å and half-metallic.

PACS numbers: 68.55.-a,75.30.-m,75.70.Ak

Half-metallic ferromagnets[1] have at $T = 0$ a finite band-gap for the majority spin bands, and a vanishing band-gap for the minority spin bands at the Fermi level, and thus can have, in principle, 100% spin-polarization.[2] There has recently been a renewed interest in them[3] for the purpose of injecting highly spin-polarized electrons into *zincblende* (ZB) semiconductor substrates. However, solids that are stable in the zincblende-type structure are normally not half-metallic[4]; conversely, known half-metallic systems possess instead the rutile structure for $CrO_2$[5], the perovskite structure[6] of e.g., $La_{1-x}Sr_xO_3$, or the Heusler structure of NiMnSb.[7] Although it is possible to grow a non-ZB film on a ZB substrate (e.g., NiAs-type MnAs on ZB-type GaAs),[8, 9] it is desirable to identify systems having geometrically coherent interfaces (i.e., same structure-type for film and substrate) so as to minimize scattering. This objective created recent interests in theoretical predictions[10–20] and experimental testing[11, 12, 21, 22] of half-metallic, metastable ZB structures made of compounds that are more stable in other structures, e.g., the NiAs-type, or MnP-type structures. These studies carried out[10–20] for MnAs, CrAs, CrSb, CrS, CrSe, and CrTe, predicted that half-metallicity is often preserved in the metastable, higher-energy ZB structure, and that there are cases where the energy difference $\Delta_{\rm bulk}$ between the equilibrium-volume ZB and non-ZB structures is quite low (0.2-1.0 eV), spurring hope that such pseudomorphic thin films could be stably grown, to the benefit of high-efficiency spin-injection devices.

In fact, however, the stability of a high-energy pseudomorphic ZB film is not decided by the energy difference $\Delta_{\rm bulk}$ between the (tri-axial) *hydrostatically deformed* ZB materials and the ground state structure, but rather by criteria considering the energetics $\Delta_{\rm epi}$ of the (bi-axial) *epitaxially deformed* film relative to the ground state (Fig. 1). Fortunately, the relevant energies can be calculated by the same first-principles methodology (but different structural configurations) used in the past to compute[10–20] the energy. The relevant energies calculated here are illustrated in Fig. 1. The solid lines indicate the normal energy *vs* volume curves for the non-ZB ground state and for the high-energy ZB phase. The energy difference between the respective minima is denoted by $\Delta_{\rm bulk}$.

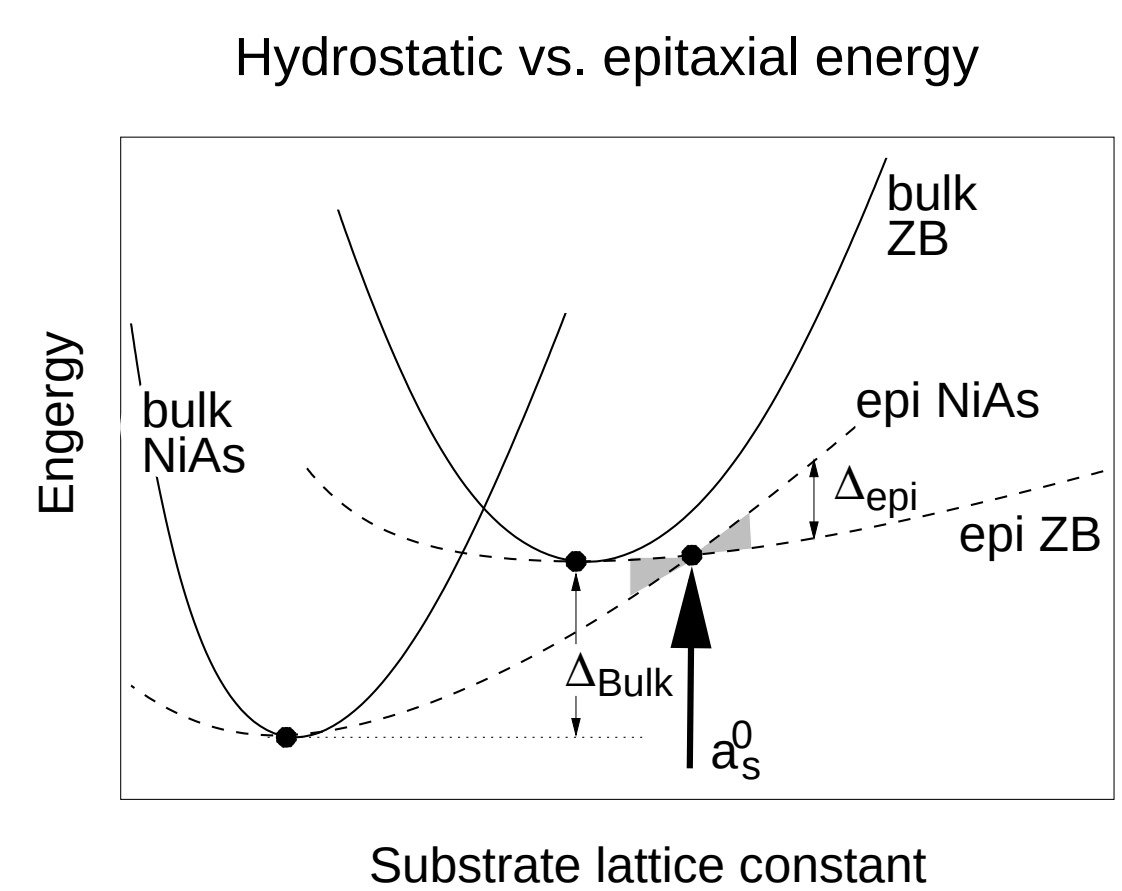

FIG. 1: The schematic plot of hydrostatic energy, tri-axial relaxed (solid lines), and epitaxial energy, bi-axial relaxed (dashed lines), as substrate lattice constant changes. Beyond $a_s^0$, the ZB epitaxial structure is more stable than the NiAs-type structure.

The dashed lines indicate the E *vs* $a_s$ epitaxial curves, corresponding to a material confined epitaxially in two in-plane substrate directions to a templet of lattice-constant $a_s$, whereas the third lattice constant and any cell-internal degree of freedom are relaxed to minimize the total energy. The epitaxial (bi-axial) curves are naturally flatter than the bulk (tri-axial) curves at the same $a_s$, since in the former case the energy is allowed to relax in one direction. The equations governing the "epitaxial softening" $E_{\rm epi}(a_s)/E_{\rm bulk}(a_s)$ are discussed in Refs [23–26]. We see from the schematic in Fig. 1 that there could exist a critical substrate lattice constant $a_s^0$ such that for $a_s > a_s^0$ the epitaxial ZB phase has a lower energy ( by $\Delta_{\rm epi}$ ) than the epitaxial NiAs structure, even though in the free-floating bulk form the NiAs structure has a lower energy. We can calculate the epitaxial energy curves for different materials as a function of $a_s$, establish the epitaxial energy difference $\Delta_{\rm epi}(a_s)$ and see if it has a zero point $\Delta_{\rm epi}(a_s = a_s^0) = 0$ or not. If it has a zero-point, we can search for an actual substrate material that has a natural lattice constant close to $a_s^0$. Furthermore, examination of the curvature of $E_{\rm epi}^{\rm ZB}(a_s)$ can tell us if this phase is mechanically stable under axial $c/a$ distortions[26] or even if it could be dynamically unstable[27]. Such calculations have to be performed both for ferromagnetic (FM) and antiferromagnetic (AFM) spin arrangement, examining if epitaxial stability comes with the desired (FM) form of magnetism. The main questions hence are: (i) Is $E_{\rm epi}(a_s)$ *vs* $a_s$ bound? (ii) If it is, does $\Delta_{\rm epi}(a_s)$ cross zero (i.e., does epi-ZB ever become lower-energy than epi-NiAs? ) (iii) If it does, is epi-ZB ferromagnetic? (iv) If it does, is there a ZB substrate material whose natural lattice constant is close to $a_s^0$ ? (this decides the thickness[23] that can be grown.)

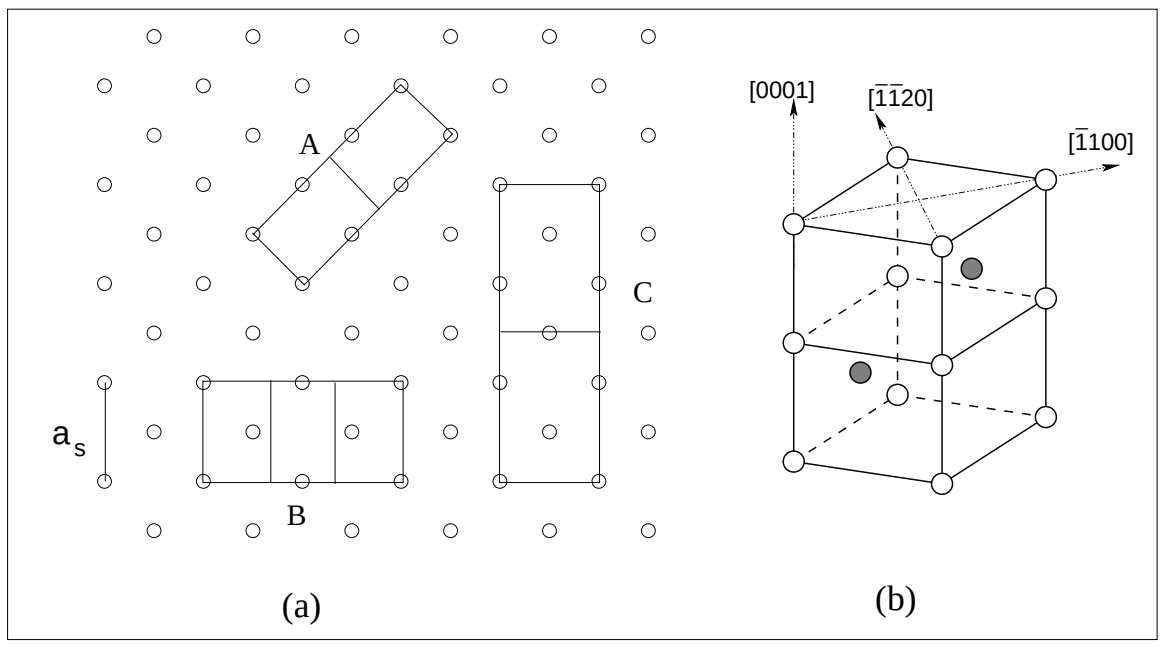

FIG. 2: Left panel shows the possible geometrical structure at the interface of zincblende (001) and NiAs-type $(\bar{1}100)$. Pattern A ($a = a_s\sqrt{2}/2$ and $c/a = 3/2$) is confirmed for NiAs-type MnAs grown on GaAs (001) substrate. The orientation of the NiAs structure is shown in right panel.

We have carried out such first-principles calculations for a number of binary systems that are thought[10–20] to be potential FM half-metallic material in their ZB form, and which were previously predicted to have rather small $\Delta_{\rm bulk}$ values, spurring hope that they can be grown pseudomorphically. Unfortunately, we found such hopes to be mostly unfounded because (i) the epitaxial energy curves $E_{\rm epi}^{\rm ZB}(a_s)$ of the ZB form are only weakly bound, especially for MnAs, CrSb (ii) $\Delta_{\rm epi}(a_s)$ does not cross zero for MnAs, CrSb, CrAs, CrTe at least up to the substrate lattice constant available from the largest lattice constant ZB substrate materials InSb or CdTe substrate ($\sim$ 6.5 Å). Thus, under epitaxial growth for a large range of $a_s$ the NiAs-type structure continues to be more stable than the ZB-type structure, just as is the case for bulk growth. (iii) $\Delta_{\rm epi}(a_s)$ does cross zero for CrS and CrSe at $a_s^0 = 5.75$ Å and 6.24 Å, respectively. However, having done so, the epitaxial ZB forms of CrS is AFM, not FM, so the system is not half-metallic. Finally, CrSe does satisfy all conditionsand could conceivably be grown as half-metallic ZB structure for $a_s > 6.2$ Å (e.g., on a $Cd_xZn_{1-x}Se$ substrate). This study shows that one would identify half-metallic epitaxial ZB structure, but that the theoretical methodology that must be used is different than hitherto practiced[10–20].

To perform epitaxial calculations for NiAs-on-ZB we need to find a relationship between the lattices of the film and the substrate such as those formulated in Ref. 8 for NaCl-on-ZB. To do so we inspect Fig. 2, which shows several possible lattice relationships between the NiAs-type lattice and ZB lattice. Indeed, some NiAs-type films, such as MnAs, have been[9, 28] experimentally known to grow along the $[\bar{1}100]$ direction, with every fourth MnAs $\{0002\}$ plane matching along the [0001] direction every sixth GaAs $\{220\}$ plane. This is pattern A shown in Fig. 2. Since $c/a$ is very close to 3/2 for all the studied NiAs-type binaries in a wide range of volumes (c.f. Fig. 3), we assume that these NiAs-type epitaxial films will grow as pattern A in Fig. 2, i.e., we fix $c/a$ at 3/2, and $a \equiv a_s\sqrt{2}/2$. In general, the *minima* of the hydrostatic $E_{\rm bulk}$ *vs* $a$ curves coincide with that of the epitaxial $E_{\rm epi}$ *vs* $a_s$ curves, e.g., see the FM NiAs-type for MnAs and CrTe in Fig. 3. However, this is not the case when the equilibrium free-floating $c/a$ value of the NiAs structure differs significantly from 3/2, as are the cases according to our calculations for CrS and CrSe (c.f. Fig. 3). It is not guaranteed that the epitaxial NiAs-type film will grow with $c/a \equiv 3/2$ (pattern A in Fig. 2) on any given substrate. However, if a NiAs-type film could be grown with a more closely lattice-matched pattern, it will only decrease the epitaxial energy of the NiAs-type structure, and thus the ZB epitaxial structure will be even more unstable with respect to the NiAs structure. We calculated the hydrostatic and epitaxial energy *vs* substrate lattice constant in both FM and AFM state for these compounds. For the AFM configuration, we let the spin in adjacent cation layers to alternate along the [110] and [0001] directions for the ZB and NiAs structures, respectively. All calculations are done with the pseudopotential momentum-space total-energy method[29] within the generalized gradient approximation of PW91 formulae[30], and the projector augmented wave (PAW) potentials, as implemented by the VASP code[31]. The charge density is obtained from the Monkhorst-Pack **k**-space integration method, using the mesh of $6\times6\times6$ for ZB and $6\times6\times4$ for the NiAs structures, with energy cut-off of 283.9 eV for CrAs, CrSb, CrSe, CrTe, and a cut off of 323.4 eV for CrS and MnAs.

Figure 3 shows the detailed results of the hydrostatic and epitaxial curves in both FM and AFM spin arrangement for these systems. The important features are summarized in Table I. We note the following: (i) MnAs, CrAs, and CrSb have large $\Delta_{\rm bulk}$, and indeed the NiAs-phase continues to be more stable than the ZB phase even under epitaxial conditions, thus, $\Delta_{\rm epi} > 0$ and no substrate lattice constant exist for which $\Delta_{\rm epi}$ crosses zero. Interestingly, CrAs is AFM in the equilibrium NiAs-type structure, but when this structure is stretched epitaxially, CrAs becomes FM. This observation is supported by the latest experiment[22] which found FM near the CrAs/GaAs interface. (ii) Although CrTe has a rather small $\Delta_{\rm bulk}$ of 0.3 eV, it too is never stabilized in the epitaxial ZB form, at least for substrate lattice constants smaller than 6.5 Å which is the largest substrate lattice constant available for binary semiconductors (CdTe, InSb). (iii) CrS has a small $\Delta_{\rm bulk}$ and its $\Delta_{\rm epi}(a_s)$ indeed does cross zero at $a_s^0 = 5.75$ Å, thus the epitaxial ZB phase can be stabilized when grown epitaxially on, e.g. CdS, CdSe, and ZnTe substrates. Unfortunately, we find (c.f. Fig. 3) that the epitaxially-stable ZB form of CrS is not ferromagnetic, thus not half-metallic. Indeed, CrS has an AFM spin arrangement in both NiAs and ZB structure. Although the ferromagnetic state of epitaxial ZB becomes stable for $a_s > 6.3$ Å, it will be difficult to grow

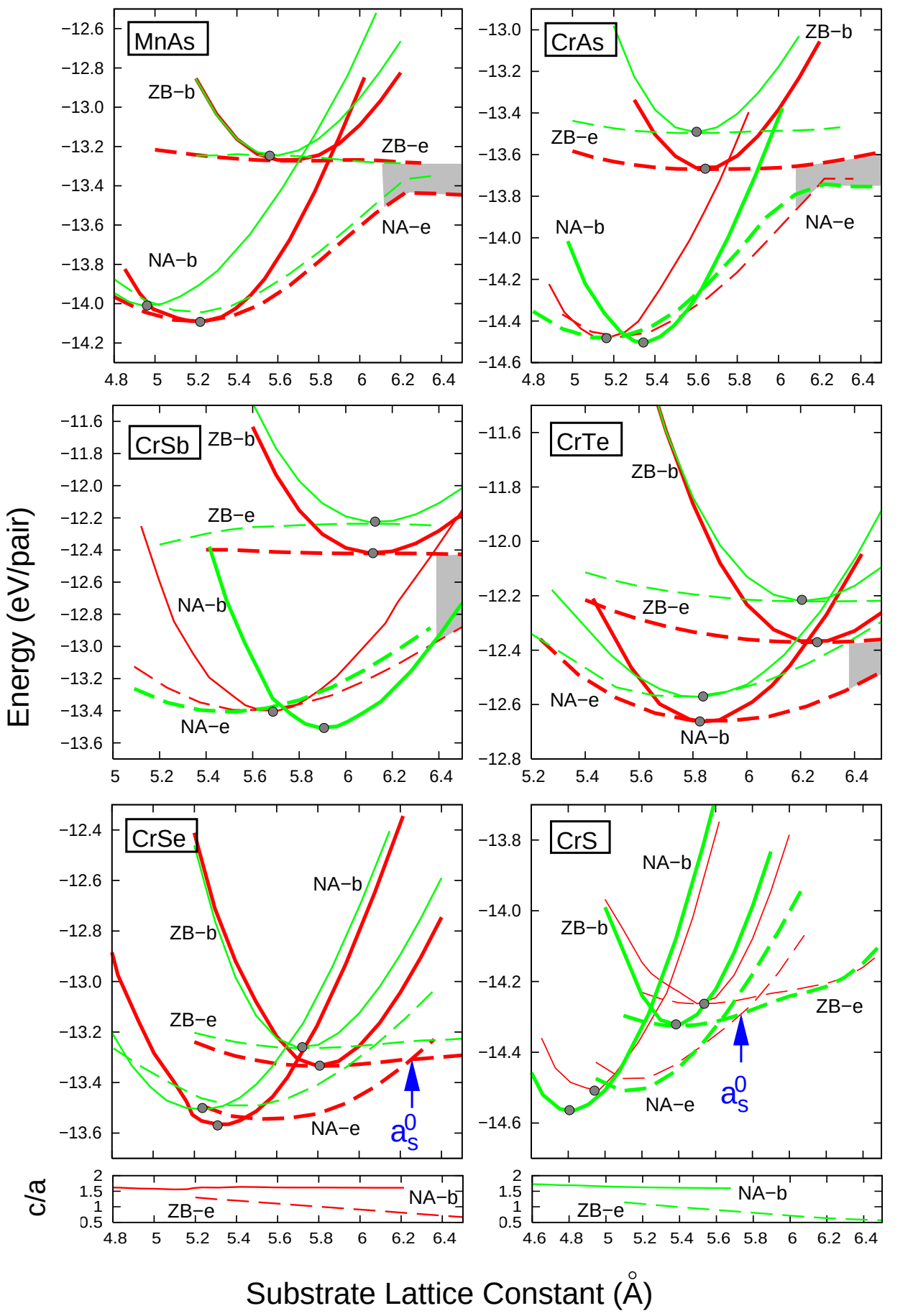

FIG. 3: (Color) The calculated hydrostatic bulk (b; solid lines) and epitaxial (e; dashed lines) total energies for both zincblende (ZB) and NiAs-type (NA) structures. Red and green lines represent ferromagnetic and antiferromagnetic spin arrangements, respectively; the stabler spin configuration is drawn in thicker line. Energies were calculated at each tenth Å of the substrate lattice constants. Vertical arrows labeled $a_s^0$ indicate that the substrate lattice constant at which epitaxial ZB structure become stabler than the NiAs counterpart in cases CrS and CrSe. Shaded areas indicate when these curves approach each other but do not cross. The bottom figures show the tetragonal $c/a$ ratios for CrSe and CrS.

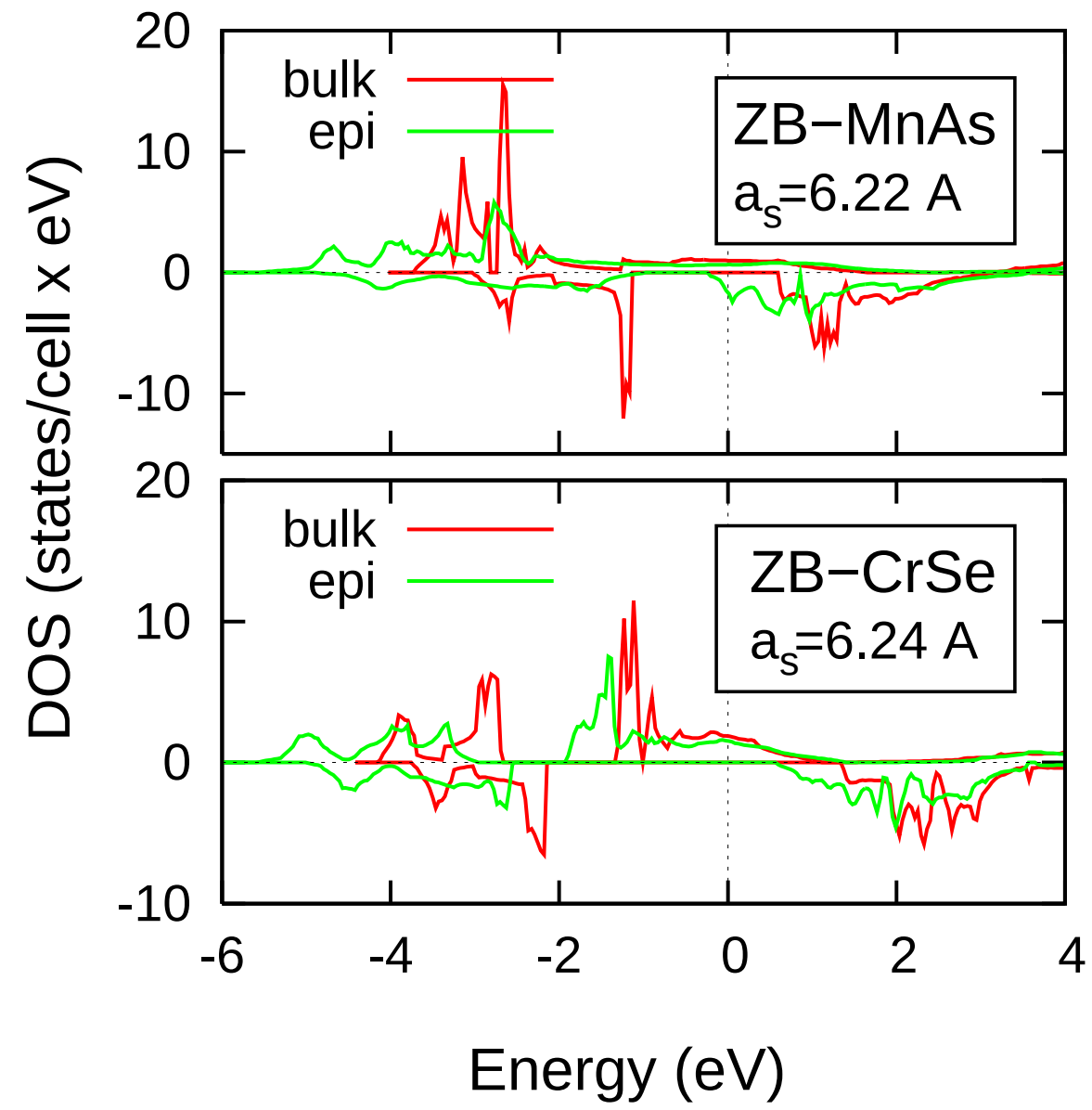

FIG. 4: (Color) Comparison of the total density of states (DOS) between epitaxial (green) and hydrostatic (red) ZB structures at 6.22 Å and 6.24 Å for MnAs and CrSe, respectively. Spin-up (down) DOS is shown on positive (negative) axis and the Fermi energy is set to zero. Note that CrSe is half-metallic in both configurations, whereas MnAs is half-metallic under hydrostatic structure, but not under epitaxial condition.

ZB CrS film due to the extremely large ($\Delta a/a \leq 17\%$) lattice mismatch making the critical thickness of the ZB film less than one monolayer[23]. (iv) Finally, CrSe has an $a_s^0$ of 6.24 Å, and thus the half-metallic[32] epitaxial ZB film could possibly be grown on a $Cd_xZn_{1-x}Te$ or $CdSe_{1-x}Te_x$ substrate whose lattice constant can be tuned to 6.2-6.3 Å using compositions of $x = 0.4$ and $x = 0.5$, respectively. We conclude that the search for half-metallic zinc-blende structure that can be grown epitaxially must follow the procedure of examining epitaxial energetics (dashed lines in Fig. 3) rather than the previously practiced[10–20] bulk energetics, and examine whether the AFM (or other) spin configurations are preferred to ferromagnetism under epitaxial conditions.

It is interesting to compare the half-metallic characteristic of bulk (=hydrostatic) *vs* epitaxial forms of the same material in the ZB structure. We see from Table I that: (i) The epitaxial films possess half-metallicity in a wider range of substrate lattice constants than their bulk phases for CrAs, CrSb, CrSe, and CrTe. (ii) However, epitaxial MnAs lacks half-metallicity even though it is half-metallic under bulk condition for $a > 5.8$ Å. (iii) ZB-CrS becomes half-metallic when $a_s > 5.5$ Å under hydrostatic condition, but under epitaxial conditions AFM dominates until $a_s > 6.3$ Å and no half-metallic character is found. The density of states (DOS) for these transition metal binaries are exemplified (Fig. 4) for CrSe and MnAs in both epitaxial and bulk phases at $a_s^0$ (or in the shaded area; c.f. Fig. 3). We see that the DOS of epitaxial phases is flatter than that of the bulk when $a_s$ is larger than the equilibrium lattice constants, indicating that the energy levels are more delocalized in epitaxial phases. (When $a_s$ is smaller than the equilibrium lattice constant, the energy levels could be more localized in the epitaxial phase.) In the case of MnAs with $a_s = 6.22$ Å (Fig. 4), the delocalized levels in conduction band of the epitaxial phase even smear to the Fermi levels for both spin channels and deprive MnAs of the half-metallicity. In the case of CrSe with $a_s = 6.24$ Å, both bulk and epitaxial phases are half-metallic even though the conduction band minimum (CBM) of spin down channel in epitaxial phase is closer to the Fermi level. When $a_s$ is smaller than the equilibrium lattice constant (e.g., $a_s = 5.50$ Å), the CBM of spin down channel in bulk CrSe is actually below the Fermi level whereas the CBM of epitaxial CrSe is above the Fermi level. This leads the bulk CrSe to a metal while keeps epitaxial CrSe as a half-metal when $a_s = 5.50$ Å (corresponding DOS is not shown in Fig. 4).

TABLE I: Calculated magnetic ground states (F=ferromagnetic, A= antiferromagnetic) of bulk and epitaxial configuration of the zincblende (ZB) and NiAs-type (NA) structure of various transition-metal binaries. We also give the calculated equilibrium lattice constants $a$ and $c/a$ ratio of the bulk phases. $\Delta_{\text{bulk}}$ denotes the amount by which the bulk NiAs structure is stabler than the bulk ZB structure at their respective equilibrium volumes (Fig. 1). We indicate whether or not the epitaxial ZB structure becomes more stable than the epitaxial NiAs structure at some critical substrate lattice constant $a_s^0$, and the lattice-constant range where epitaxial ZB is half-metallic (HF).

| | NiAs-type | | ZB-type | | $\Delta_{\text{bulk}}$ (eV) | Epi NiAs | ZB HM range | |
|---|---|---|---|---|---|---|---|---|
| | Bulk | Epi | Bulk | Epi | | $\to$ Epi ZB? | Bulk | Epi |
| | $a$(Å), $c/a$ | | $a$(Å) | | | $a_s^0$ (Å) | (Å) | (Å) |
| MnAs | F$\sim$A<br>3.68, 1.496 | F | F$\sim$A<br>5.67 | F$\sim$A | 0.82 | No | $>5.8$ | – |
| CrAs | F$\sim$A<br>3.78, 1.390 | F$\sim$A | F<br>5.64 | F | 0.84 | No | $>5.6$ | $>4.9$ |
| CrSb | F$\sim$A<br>4.18, 1.276 | F$\sim$A | F<br>6.11 | F | 1.08 | No | $>5.8$ | $>5.6$ |
| CrS | F$\sim$A<br>3.41, 1.691 | F$\sim$A | F$\sim$A<br>5.37 | A | 0.24 | Yes<br>$a_s^0=5.75$ | $>5.5$ | – |
| CrSe | F$\sim$A<br>3.76, 1.616 | F$\sim$A | F$\sim$A<br>5.77 | F | 0.23 | Yes<br>$a_s^0=6.24$ | $>5.7$ | $>4.9$ |
| CrTe | F<br>4.13, 1.526 | F | F<br>6.24 | F | 0.30 | No | $>6.0$ | $>5.1$ |

In summary, we show that the energetic proximity of the *bulk* total-energy minima $\Delta_{\text{bulk}}$ of the stable and metastable crystal structure types is not sufficient to indicate that the latter could be stabilized as thin-film. Instead, one has to examine the corresponding *epitaxial* total-energy difference curve $\Delta_{\text{epi}}(a_s)$ for various substrate lattice constants in search for $a_s = a_s^0$ which reverts the stability of the competing phases. We illustrate this procedure for a few binary transition metal compounds that were proposed as half-metallic ZB structures, finding that MnAs, CrSb, CrAs, and CrTe have $\Delta_{\text{epi}} > 0$, so under epitaxial conditions they continue to be stabler in the NiAs-type structure than in ZB, just as is the case in bulk forms. CrS does have $\Delta_{\text{epi}} < 0$, but is not ferromagnetic under those conditions. Finally CrSe has $\Delta_{\text{epi}} < 0$ for $a_s > 6.2$ Å and *is* half-metallic in epitaxial forms. It might be grown on the $Cd_xZn_{1-x}Se$ substrate with $x = 0.4$.

This work was supported by the Defense Advanced Research Projects Agency (DARPA).